# Coulomb Blockade Thermometry on a Wide Temperature Range

O.M. Hahtela[1,†], A. Kemppinen[1], J. Lehtinen[1], A.J. Manninen[1,*], E. Mykkänen[1], M. Prunnila[1], N. Yurttagül[1], F. Blanchet[2], M. Gramich[2,††], B. Karimi[2], E.T. Mannila[2], J. Muhojoki[2], J.T. Peltonen[2], and J.P. Pekola[2]

[1]VTT Technical Research Centre of Finland Ltd, P.O. Box 1000, FI-02044 VTT, Espoo, Finland
[2]QTF Centre of Excellence, Department of Applied Physics, Aalto University, Aalto, FI-00076, Finland
[*]E-mail: antti.manninen@vtt.fi
[†]Present address: Vaisala Oyj, Vanha Nurmijärventie 21, FI-01670 Vantaa, Finland
[††]Present address: Department of Applied Sciences and Mechatronics, MUAS, Lothstr. 34, 80335 Munich, Germany

*Abstract* — **The Coulomb Blockade Thermometer (CBT) is a primary thermometer for cryogenic temperatures, with demonstrated operation from below 1 mK up to 60 K. Its performance as a primary thermometer has been verified at temperatures from 20 mK to 200 mK at uncertainty level below 1 % ($k$ = 2). In a new project, our aim is to extend the metrologically verified temperature range of the primary CBT up to 25 K. We also demonstrate close-to-ideal operation of a CBT with only two tunnel junctions when the device is embedded in a low-impedance environment.**

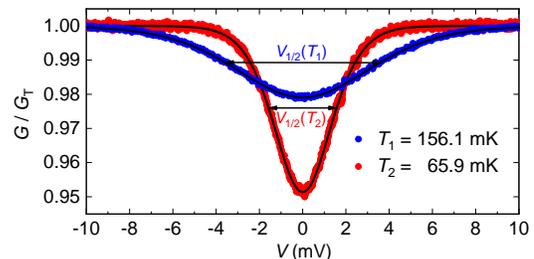

Fig. 1. Scaled dynamic conductance of a CBT with $N$ = 99 and $E_C/k \approx$ 10 mK at two temperatures, $T_1$ = 156.1 mK (blue dots) and $T_2$ = 65.9 mK (red dots) with fits to the theory (black lines).

## I. Introduction

Since 20 May 2019, the definition of the SI unit kelvin has been based on the fixed value of the Boltzmann constant $k$. The traceability for temperature measurements can be provided directly by primary thermometry, without reference to defined temperature scales such as ITS-90 or PLTS-2000. Research on practical primary methods for realization and dissemination of temperature in terms of the redefined kelvin has been especially active at cryogenic thermometry. One of the most advanced methods is the Coulomb Blockade Thermometer (CBT) [1], which has already been demonstrated to operate as a primary thermometer from below 1 mK up to 60 K or above.

## II. Operation Principle and Recent Progress

The CBT is a purely electronic, solid-state thermometer based on single-electron charging effects. The dynamic conductance $G = dI/dV$ of an array of $N$ micro- or nanosize tunnel junctions connected in series with normal-metal islands in between has a dip as a function of bias voltage at $V = 0$, with depth proportional to $E_C/kT$ and full width at half minimum [1]

$$V_{1/2}(T) \cong 5.439\,NkT/e \quad (1)$$

(see Fig. 1). Here $T$ is the temperature of electrons in the array, $k$ is the Boltzmann constant, $e$ is the elementary charge, $E_C = [(N-1)/N]e^2/C$ is the characteristic charging energy of the system, and $C$ is the total capacitance of one island. Since $V_{1/2}$ depends linearly on $T$ but not on $E_C$ or any other device-dependent parameters (except for $N$), it can be used for primary thermometry. Equation (1) is strictly valid in the weak Coulomb blockade regime ($kT \gg E_C$), but the operation range of the CBT can be extended to the intermediate Coulomb blockade regime ($kT \approx E_C$), still conserving its primary nature [2]. Other requirements for the validity of Eq. (1) are that the tunnel resistances exceed the resistance quantum $R_Q \equiv h/e^2 \approx$ 26 kΩ, the array is homogeneous, and $N$ is large enough (typically 30 – 100) to suppress errors caused by the electromagnetic environment of the array. Several series-connected chains are connected in parallel to decrease the impedance to a more convenient level.

The original experiments and most of the later studies of the CBT have focused on temperatures below 4.2 K, and so far the CBT has found its most important applications at temperatures below 500 mK. According to a careful uncertainty analysis, the relative combined uncertainty of the CBT as a primary thermometer is below 1 % ($k$ = 2) at temperatures from 20 mK to 200 mK [3], [4]. This was verified in a comparison between the CBT and two other primary thermometers (the current sensing noise thermometer and the magnetic field fluctuation thermometer) and the PLTS-2000 temperature scale [5].

At temperatures below about 20 mK, the weak electron-phonon coupling with $T^5$ temperature dependence starts to limit the feasibility of the CBT in practical thermometry. $V_{1/2}$ of Eq. (1) depends on the electron temperature of the CBT, but typically the measured object is thermally coupled with the CBT via the phonon system. By increasing the volume of the metallic islands to enhance the electron-phonon coupling, temperatures below 4 mK have been measured by the CBT [6], [7]. Even lower electron temperatures, down to below 0.5 mK, have been measured by cooling the metallic islands of the CBT directly by nuclear demagnetization [7] – [10].

On the high-temperature end of the operation range, the dimensions of the CBT must be very small to increase $E_C$ to a value that produces a deep enough conductance dip for accurate measurements. Fabrication of a homogeneous array of tunnel junctions with dimensions 50 nm or below is a challenging task. Another problem at high temperatures is the non-constant background conductance caused by the finite tunnel barrier height. However, operation of the CBT with about 1 % precision has already been demonstrated up to $T \approx 60$ K [11].

## III. TOWARDS TRACEABLE CBT THERMOMETRY UP TO 25 K

In September 2019, 14 European NMIs and 7 other partners started a joint research project Real-K (Realization of the redefined kelvin) in the EMPIR programme. One objective is to develop primary thermometry for temperatures between 1 K and 25 K, and the CBT is one of the developed methods.

Two approaches for fabricating the new, traceable CBTs are in progress. One of them is the shadow evaporation method using a thin e-beam patterned germanium film as a mask [11]. We have recently developed the fabrication process further and have been able to decrease the e-beam patterning time by a factor of 20 from the earlier process [11], and Fig. 2 shows a scanning electron micrograph of the first CBT device fabricated with the new process. The other approach for CBT fabrication is to use the sidewall-passivating spacer structure (SWAPS) process, which was originally developed for fabricating sub-$\mu$m Josephson junctions starting from a Nb-AlO$_x$-Nb trilayer [12]. The dynamic conductance $G(V)$ of the CBT is determined by modulating the DC bias with a small AC voltage and measuring the AC voltage and current of the CBT with two lock-in amplifiers or sampling precision multimeters [4]. In the end of the project, comparison measurements between the CBTs and other developed primary thermometers will be performed.

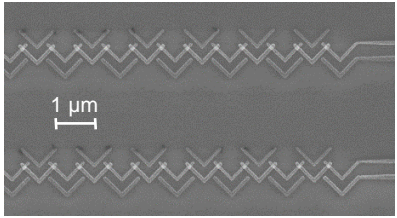

Fig. 2. One end of two series-connected chains of about $50 \times 50$ nm$^2$ tunnel junctions of a CBT for thermometry between 1 K and 25 K. The device has 50 parallel chains of 100 series-connected tunnel junctions.

## IV. TWO-JUNCTION CBT

Errors caused by the electromagnetic environment of the CBT are suppressed by connecting a large number of tunnel junctions in series. Another approach would be to use a CBT with only two tunnel junctions in series and to embed it in a very low-impedance environment. In the limit of vanishing impedance, a two-junction CBT is expected to follow Eq. (1) again [13]. We have obtained promising results in experiments with a two-junction CBT in a low-impedance environment realized by on-chip capacitors of about 50 pF. As Fig. 3 shows, the relative deviation between the temperature readings of a two-junction CBT and a conventional CBT with $N = 100$ is within ±5 % at temperatures between 25 mK and 75 mK.

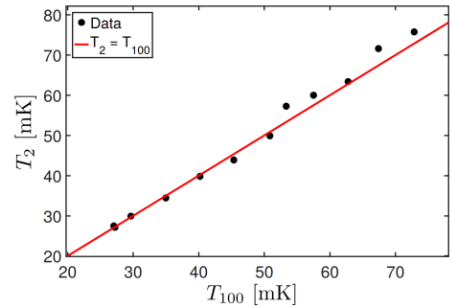

Fig. 3. Temperature readings of a two-junction CBT ($T_2$) and a 100-junction CBT ($T_{100}$). The red line corresponds to $T_2 = T_{100}$.

ACKNOWLEDGEMENT

Project Real-K has received funding from the European Metrology Programme for Innovation and Research (EMPIR) co-financed by the Participating States and from the European Union's Horizon 2020 research and innovation programme.

REFERENCES

[1] J.P. Pekola, K.P. Hirvi, J.P. Kauppinen, and M.A. Paalanen, "Thermometry by arrays of tunnel junctions," *Phys. Rev. Lett.*, vol. 73, no. 21, pp. 2903 – 2906, November 1994.
[2] A.V. Feshchenko *et al.*, "Primary thermometry in the intermediate Coulomb blockade regime," *J. Low Temp. Phys.*, vol. 173, pp. 36 – 44, April 2013.
[3] O.M. Hahtela *et al.*, "Investigation of uncertainty components in Coulomb blockade thermometry," *AIP Conf. Proc.*, vol. 1552, pp. 142 – 147, September 2013.
[4] O. Hahtela *et al.*, "Traceable Coulomb blockade thermometry," *Metrologia*, vol. 54, no. 1, pp. 69 – 76, December 2016.
[5] J. Engert *et al.*, "New evaluation of $T - T_{2000}$ from 0.02 K to 1 K by independent thermodynamic methods," *Int. J. Thermophys.*, vol. 37, art. 125, October 2016.
[6] D.I. Bradley *et al.*, "Nanoelectronic primary thermometry below 4 mK," *Nature Comm.*, vol. 7, art. 10455, January 2016.
[7] M. Palma *et al.*, "On-and-off chip cooling of a Coulomb blockade thermometer down to 2.8 mK," *Appl. Phys. Lett.*, vol. 111, no. 25, art. 253105, December 2017.
[8] D.I. Bradley *et al.*, "On-chip magnetic cooling of a nanoelectronic device," *Sci. Rep.*, vol. 7, art. 45566, April 2017.
[9] N. Yurttagül, M. Sarsby, and A. Geresdi, "Indium as a high-cooling-power nuclear refrigerant for quantum nanoelectronics," *Phys. Rev. Applied*, vol. 12, no. 1, art. 011005, July 2019.
[10] M. Sarsby, N. Yurttagül, and A. Geresdi, "500 microkelvin nanoelectronics," arXiv:1903.01388, March 2019.
[11] M. Meschke, A. Kemppinen, and J.P. Pekola, "Accurate Coulomb blockade thermometry up to 60 kelvin," *Phil. Trans. R. Soc. A*, vol. 374, no. 2064, art. 20150052, March 2016.
[12] L. Grönberg *et al.*, "Side-wall spacer passivated sub-µm Josephson junction fabrication process," *Supercond. Sci. Technol.*, vol. 30, no. 12, art. 125016, November 2017.
[13] J.P. Pekola, T. Holmqvist, and M. Meschke, "Primary tunnel junction thermometry," *Phys. Rev. Lett.*, vol. 101, no. 20, art. 206801, November 2008.